\documentclass{pasj01}
\usepackage{url}
\usepackage{lineno}
\usepackage{comment}
\Received{$\langle$reception date$\rangle$}
\Accepted{$\langle$acception date$\rangle$}
\Published{$\langle$publication date$\rangle$}

\begin{document}
\title{The Centimeter to Submillimeter Broad Band Radio Spectrum of the Central Compact Component in A Nearby Type-II Seyfert Galaxy NGC~1068}
\author{Tomonari Michiyama\altaffilmark{1,2,3}}%
\altaffiltext{1}{Department of Earth and Space Science, Osaka University 1-1 Machikaneyama, Toyonaka, Osaka 560-0043, Japan}
\altaffiltext{2}{National Astronomical Observatory of Japan, National Institutes of Natural Sciences, 2-21-1 Osawa, Mitaka, Tokyo, 181-8588}
\altaffiltext{3}{Faculty of Welfare and Information, Shunan University, 43-4-2 Gakuendai, Shunan, Yamaguchi, 745-8566, Japan}

\author{Yoshiyuki Inoue\altaffilmark{1,4,5}}%
\altaffiltext{4}{Interdisciplinary Theoretical \& Mathematical Science Program (iTHEMS), RIKEN, 2-1 Hirosawa, Saitama, 351-0198, Japan}
\altaffiltext{5}{Kavli Institute for the Physics and Mathematics of the Universe (WPI), UTIAS, The University of Tokyo, 5-1-5 Kashiwanoha, Kashiwa, Chiba 277-8583, Japan}

\author{Akihiro Doi\altaffilmark{6,7}}%
\altaffiltext{6}{The Institute of Space and Astronautical Science, Japan Aerospace Exploration Agency, 3-1-1 Yoshinodai, Chuou-ku, Sagamihara, Kanagawa 252-5210, Japan}
\altaffiltext{7}{Department of Space and Astronautical Science, SOKENDAI, 3-1-1 Yoshinodai, Chuou-ku, Sagamihara, Kanagawa 252-5210, Japan}

\email{t.michiyama.astr@gmail.com}
\maketitle

\begin{abstract}
We analyze all the available Atacama Large Millimeter / submillimeter Array archival data of the nearby Type-II Seyfert galaxy NGC~1068, including new 100\,GHz data with the angular resolution of 0\farcs05, which was not included in previous continuum spectral analysis.
By combining with the literature data based on the Very Large Array, we investigate the broadband radio continuum spectrum of the central $\lesssim7$~pc region of NGC~1068.
We found that the flux density is between $\approx$10--20\,mJy at 5--700\,GHz. 
Due to the inability of the model in previous studies to account for the newly added 100\,GHz data point, we proceeded to update the models and make the necessary adjustments to the parameters.
One possible interpretation of this broadband radio spectrum is a combination of emission from the jet base, the dusty torus, and the compact X-raying corona with the magnetic field strength of $\approx20$\,G on scales of $\approx30$ Schwarzschild radii from the central black hole. In order to firmly identify the compact corona by omitting any other possible extended components (e.g., free-free emission from ionized gas around), high-resolution/sensitivity observations achieved by next-generation interferometers will be necessary. 
\end{abstract}

\section{Introduction}
The nearby spiral galaxy NGC~1068 located at the distance of $D_{\rm L}=13.97\pm2.10$~Mpc \citep{Anand2021MNRAS.501.3621A} is considered as the archetype for type-II Seyfert in unified schemes, showing broad-line emission in polarized optical continuum \citep{Miller1983ApJ...271L...7M, Antonucci1985ApJ...297..621A}.
The most notable observations of NGC~1068 in recent years is the tentative TeV neutrino detection by the IceCube observatory \citep{IceCube, IceCube2022Sci...378..538I}, and a pressing issue is to solve multi-messenger puzzle \citep{Inoue_2020,Murase2020PhRvL.125a1101M,Kheirandish2021ApJ...922...45K,Eichman2022ApJ...939...43E,Inoue2022arXiv220702097I,Michiyama2022ApJ...936L...1M}.

In the centimeter-wave band, the relativistic kpc scale jet is the dominant source for the entire emission of NGC~1068 and multiple bright components exist \citep{Wilson1987ApJ...319..105W}. Among those components, the nuclear region has been identified and labeled as S1 \citep{Gallimore_1996_2,Gallimore_1996_1, Anand2021MNRAS.501.3621A}  by high-resolution interferometric observations (Very Large Array; VLA and Multi-Element Radio Linked Interferometer Network; MERLIN).
However, the origin of the radio continuum at S1 is still under debate.

\citet{Gallimore2004ApJ...613..794G} reported the detection of S1 at 5~GHz, but not at 1.4~GHz by Very Long Baseline Array (VLBA). They argued that thermal free-free emission from an X-ray heated corona or ionized accretion disk wind may be the origin because the average brightness temperature is too low for synchrotron self-absorption to explain the 1.4\,GHz non-detection.
However, a probable detection of linear polarization at 22\,GHz \citep{Gallimore_1996_2} indicates that some of the radio emissions may be of synchrotron origin. Later, \citet{Cotton2008A&A...477..517C} added the measurements at 43\,GHz, and showed the flat spectrum from 5~GHz which indicates thermal origin for S1 rather than synchrotron origin.

Recently, using Atacama Large Millimeter/submillimeter Array (ALMA), the millimeter continuum emission of S1 has been investigated \citep{Inoue_2020}. They proposed the millimeter excess ($>200$\,GHz) with respect to the synchrotron self-absorption spectrum originating from the corona. On the other hand, \citet{Baskin2021MNRAS.508..680B} interpreted the millimeter components by the free-free emission from the gas just outside the broad-line region through the radiation pressure compression mechanism.

In this paper, to update the analysis by \citet{Inoue_2020}, we obtain the representative spectral energy distribution (SED) including the new $\approx100$\,GHz ALMA observations and investigate the origin of the centimeter/submillimeter continuum emission around the vicinity of active supermassive black holes.
Because the measurements are based on radio interferometers, we investigate the flux rise/fall seen at the centimeter/millimeter range with extra caution of the synthesized beam, maximum recovery scales, and two-dimensional fitting processes.
This paper is structured as follows: Section~\ref{sec:SED} explains the ALMA and VLA data we used. In Section~\ref{sec:discussion}, we investigate the possible origin of the SED.

\section{Representative SED} \label{sec:SED}
\begin{figure*}[!htbp]
\begin{center}
\includegraphics[angle=0,scale=1.0]{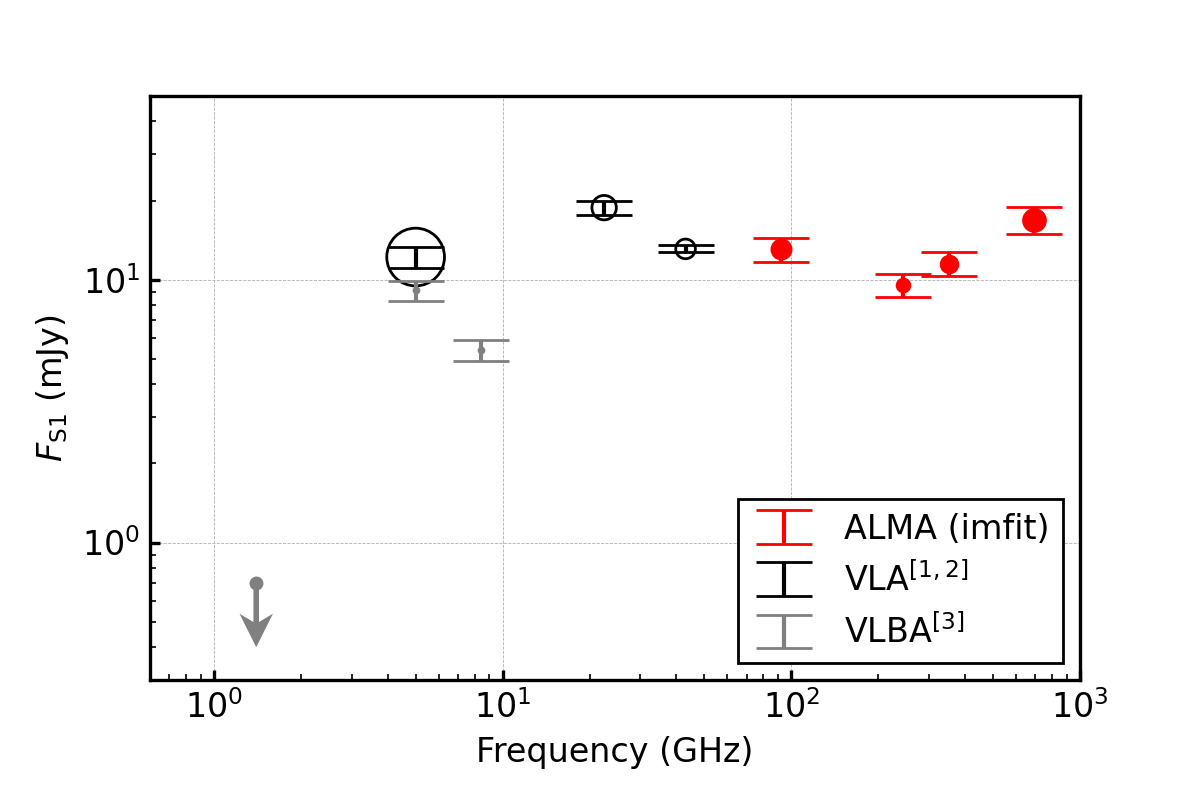}
\caption{\raggedright Representative SED around the core (S1) of NGC~1068. All measurements represent the flux in the component that being a 2D Gaussian fit falls around S1. The red ALMA points are obtained in Section~\ref{sec:ALMA}. The black open circles represent measurements by VLA (\cite{Gallimore_1996_1,Cotton2008A&A...477..517C}, see Section~\ref{sec:lit_VLA}).
The dashed grey open circles represent measurements by VLBA (\cite{Gallimore2004ApJ...613..794G}, see Section~\ref{sec:lit_VLBA}).
The arrow at 1.4\,GHz indicates the non-detection upper limit.
The size of the circle indicates the synthesized beam (the geometric mean of the major/minor axis of an ellipse) for each image.
\label{fig:rep_SED}}
\end{center} 
\end{figure*}

This section explains how the representative radio spectrum (figure~\ref{fig:rep_SED}; 1-1000\,GHz) is obtained.
For ALMA data ($>90$\,GHz), we make the representative map for each ALMA receiver (Band~3, 6, 7, 9) using archival data. For the VLA and VLBA measurements, we use the literature data. The details are explained in the following sub-sections.

\subsection{ALMA Archival Data} \label{sec:ALMA}
\begin{figure*}[!htbp]
\begin{center}
\includegraphics[angle=0,scale=0.22]{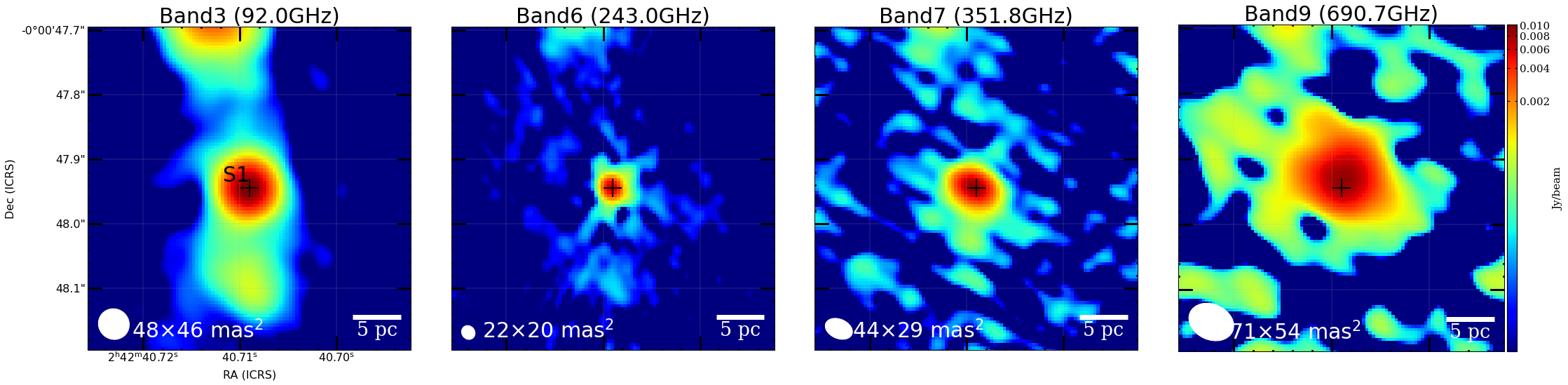}
\caption{\raggedright Representative images combining all the archival data. The white ellipse at the bottom left corner represents the synthesized beam, and the white bar at the bottom right corner represents the physical scale bar.
The S1 is at 
$(\alpha,\delta)_{\rm ICRS}=$ (\timeform{02h42m40s.70901}, $-$\timeform{00D00'47''.9448}) \citep{BH_position}.
We note that the optically-defined FK5 frame differs from ICRS by a maximum of 31.57 milliarcseconds (mas) and the CASA J2000 frame\footnote{J2000 is also used as a name to label the ``epoch", but here the term of J2000 is used to mention the ``frame" in Common Astronomy Software Applications package ({\tt CASA}; \cite{McMullin_2007}. The conversion method is explained in \url{https://safe.nrao.edu/wiki/bin/view/ALMA/ICRSToJ2000}} differs from ICRS by 23.15 mas. Therefore, the coordinates of S1 are
$(\alpha,\delta)_{\rm FK5,J2000}=$ (\timeform{02h42m40s.71053}, $-$\timeform{00D00'47''.9509}) and 
$(\alpha,\delta)_{\rm CASAJ2000}=$ (\timeform{02h42m40s.70998}, $-$\timeform{00D00'47''.961848}), respectively.
\label{fig:rep}}
\end{center} 
\end{figure*}

In order to avoid any contamination from the host galaxy, high-resolution data is necessary. Table~\ref{tab:ALMA} is a summary of ALMA projects that achieved the synthesized beamsize of $<0\farcs1$ corresponding to $<7$~pc in the physical scale.
For all data from the archive, the data calibration and imaging processes were performed using {\tt CASA} \citep{McMullin_2007,CASA2022arXiv221002276T}.
We use the reduction script provided by the observatory to restore the calibrated measurement set (MS) for each EB using the specified {\tt CASA} version. For imaging, we use {\tt CASA} version of 6.1.1.15.
We make the representative images for each receiver (Band~3,6,7, and 9) by combining all the available measurement sets and using the data sampled in uv-distance of larger than $>300$~k$\lambda$ (figure~\ref{fig:rep}).
In each project, we avoided the spectral window (spw) in which line observations are targetted. In addition, we made continuum maps for every spw, confirming no line contamination in our representative map.
The maps were produced using the {\tt tclean} task in {\tt CASA} with Briggs weighting (robust = 0.5) and a pixel size of 5\,mas. The clean masks were selected by 100 pix around position S1. This simple clean mask enables us to produce representative images from a large data set in realistic computation time.
\begin{table}
\tbl{Summary of the ALMA archival data}{%
\begin{tabular}{lll}
\hline
Project Code & P.I. & Band \\
(1) & (2) & (3) \\
\hline
2013.1.00014.S & Elitzur, M & 9\\
2013.1.00055.S & Garcia-Burillo, S & 9\\
2016.1.00052.S & Imanishi, M & 6\\
2016.1.00232.S & Garcia-Burillo, S & 6, 7\\
2017.1.01666.S & Gallimore, J & 6 \\
2018.1.00037.S & Imanishi, M & 6 \\
2018.1.01135.S & Wang, J & 3 \\
2018.A.00038.S & Maeda, K & 3\\
\hline
\end{tabular}}
\label{tab:ALMA}
\end{table}

For Band~3 data, we use only 2018.1.01135.S as a representative image because combining the data of 2018.1.01135.S and 2018.A.00038.S is difficult due to different phase centers and the rich integration time was assigned in 2018.1.01135.S for line observations.
We note that the project 2016.1.00176.S met our selection criteria (achieved angular resolution is $<0\farcs$1). However, we do not use this project data, since several issues are known to exist during QA2 processes regarding polarization analysis based on the reports. Using the data of 2016.1.00176.S, \citet{Lopez-Rodriguez2020ApJ...893...33L} have measured the polarization claiming the polar dust around the nucleus. Arguments about nuclear polarization are beyond the scope of this paper.

The peak flux density ($F_{\rm peak}$) around S1 and the noise level of the map ($\sigma_{\rm rms}$) are shown in table~\ref{tab:rep} and the synthesized beamsize ($b_{\rm maj}$ and $b_{\rm min}$) is shown in figure~\ref{fig:rep} and table~\ref{tab:imfit}.
We also checked the literature measuring the continuum flux density with the data which achieve the synthesized beam of $<0\farcs1$ \citep{Garc2016ApJ...823L..12G,Impellizzeri2019ApJ...884L..28I,Inoue_2020}.
Because each literature adopts different analysis procedures to measure the flux density, it is difficult to construct the radio spectrum based on the literature data. 
In order to reduce the systematic uncertainties among literature values, the ALMA imaging analysis and flux measurements are performed by ourselves for each data in this paper.

\begin{table}
\tbl{The information of representative images.}{%
\begin{tabular}{ccc}
\hline
frequency & $F_{\rm peak}$ & $\sigma_{\rm rms}$ \\
GHz & mJy~beam$^{-1}$ & mJy~beam$^{-1}$ \\
(1) & (2) & (3)\\
\hline
92  & 11.1 & 0.02 \\
243  & 6.7 & 0.01  \\
346  & 7.1 & 0.04 \\
690  & 9.3 & 0.35 \\
\hline
\end{tabular}}
\label{tab:rep}
\begin{tabnote}
(1) observed frequency, (2) the peak flux (i.e., unresolved flux) around S1, (3) The sensitivity of the map.
\end{tabnote}
\end{table}

\begin{table*}
\tbl{The measurements of the representative SED.}{%
\begin{tabular}{ccccccccc}
\hline
telescope & frequency & $b_{\rm maj}\times b_{\rm min}$ & $F_{\rm S1}$ & $\theta_{\rm maj}$ & $\theta_{\rm min}$ & PA & ref.\\
& GHz & mas$^2$ & mJy & mas & mas & deg & \\
(1) & (2) & (3) & (4) & (5) & (6) & (7) \\
\hline
ALMA & 92 & 49$\times$47 & 13.0$\pm$1.3 & 25.5$\pm$2.0 & 18.1$\pm$2.6 & -4$\pm$14 & this work \\
ALMA & 243 & 22$\times$20 & 9.6$\pm$1.0 & 15.5$\pm$0.5 & 13.6$\pm$0.5 & 5$\pm$12 & this work \\
ALMA & 352 & 44$\times$30 & 11.5$\pm$1.2 & 21.7$\pm$1.6 & 18.3$\pm$2.1 & 18$\pm$24 & this work \\
ALMA & 691 & 71$\times$54 & 16.9$\pm$2.0 & 62.5$\pm$8.7 & 59.3$\pm$10.1 & 0$\pm$90 & this work \\
VLA & 5 & $490\times380$ & 12.2$\pm$1.1 & $72.0^{78.9}_{64.7}$ & $33.2^{41.2}_{23.3}$ & $26.5\pm4.4$ & Gallimore+96 \\
VLA & 22 & $82\times73$ & 18.8$\pm$1.1 & $42.9^{48.9}_{36.1}$ & $5.6^{21.4}_{0}$ & $-13.4\pm2.8$ & Gallimore+96 \\
VLA & 43 & $50\times50$ & 13.1$\pm$0.4 & $33.4\pm2$ & $26.4\pm2.5$ & $-16\pm13$ & Cotton+08 \\
VLBA & 1.4 & $16.0\times7.6$ & $<0.06$ $(<0.7)$ &--&--&--& Gallimore+04\\
VLBA & 5.0 & $4.7\times2.1$ & 9.1$\pm$0.8 & $16.6\pm0.2$ & $11.2\pm2.4$ & $-75.5\pm3.9$ & Gallimore+04 \\
VLBA & 8.4 & $4.7\times2.1$ & 5.4$\pm$0.5 & $11.0\pm0.4$ & $3.7\pm0.8$ & $-71.9\pm1.4$ & Gallimore+04 \\
\hline
\end{tabular}}
\label{tab:imfit}
\begin{tabnote}
(1) telescope name, (2) observed frequency, (3) synthesized beam size, (4) the flux of the component around S1 investigated by {\tt imfit}, (5) beam-deconvolved major axis length (FWHM), (6) minor axis length, and (7) position angle. The position angle is defined by $[-90, 90]$ deg where the PA=0\,deg represents north-south direction. For $F_{\rm S1}$ of VLBA 1.4~GHz, we show the upper limits after masking analysis in parentheses.
\end{tabnote}
\end{table*}

\subsection{VLA literature data}\label{sec:lit_VLA}
The 5\,GHz and 22\,GHz continuum flux density is measured by VLA \citep{Gallimore_1996_1}. We note that the 5\,GHz observation did not achieve $<0\farcs1$ angular resolution. However, the flux density at 5\,GHz is important because of the upper limits observed in low-resolution data. In Section A.2 of \citet{Gallimore_1996_1}, the detailed Gaussian fitting procedure is explained.
The results of VLA flux measurement is shown in table~4 of \citet{Gallimore_1996_1} and these results are directly copied to table~\ref{tab:imfit}.
The 43\,GHz 50\,mas resolution image obtained by VLA is presented in \citet{Cotton2008A&A...477..517C}. 
The results of the  2D elliptical gaussian fitting are shown in table~1 of \citet{Cotton2008A&A...477..517C}.
We directly copy their results into our table~\ref{tab:imfit}.

\subsection{VLBA literature data}\label{sec:lit_VLBA}
The VLBA 1.4\,GHz, 5.0\,GHz, and 8.4\,GHz images were presented in \citet{Gallimore2004ApJ...613..794G}. The 1.4\,GHz continuum is not detected at S1. This upper limit is significant compared to the spectrum $>10$\,GHz.
For 5.0\,GHz, the standard imaging processes are performed. However, for 8.4\,GHz, some specific imaging procedures are implemented to recover extended emission.
They state 
{\it ``We improved the recovery of extended emission in these data by employing the same DIFMAP self-calibration and multiresolution CLEAN deconvolution"}.
In table~2 of \citet{Gallimore2004ApJ...613..794G}, the results of image moment analysis are shown. 
We directly copy the flux shown in table~2 of  \citet{Gallimore2004ApJ...613..794G} into our table~\ref{tab:imfit}.
We do not use table~3 of \citet{Gallimore2004ApJ...613..794G} in which additional masking and taper analysis are performed to measure the spectral index.

\subsection{SED} \label{sec:SED_sub}
We measure the flux of S1 ($F_{\rm S1}$) and beam deconvolved source size (represented by $\theta_{\rm maj}$, $\theta_{\rm min}$, and position angle: PA) using {\tt imfit} task in {\tt CASA} by ourselves.
The results are shown in table~\ref{tab:imfit}.
The error ($\sigma_\mathrm{imfit}$) is estimated using the imfit task, which is based on the fitting error. However, systematic errors might be larger than $\sigma_\mathrm{imfit}$. Therefore, we assume the $10\%$ systematic error as
\begin{equation}
\sigma=\sqrt{\sigma_\mathrm{imfit}^2+{(0.1F_{\rm S1})}^2}.
\end{equation}
We use images whose minimum baseline used for imaging is the same among Band~3, 6, 7, and 9 ($>300$\,k$\lambda$). 
We do not add any other corrections (i.e. restoring the synthesized beam, masking the aperture region, and aperture photometry).
The fluxes of Bands~6,7, and 9 in this paper differ from those of \citet{Inoue_2020}, since we use different data sets and flux measurement methods are not exactly the same.
For the VLA data, the flux density measured by the 2D Gaussian fitting in the literature is used.
For the VLBA data, we use 2D Gaussian fitting values without masking analysis for 5\,GHz and 8.4\,GHz
whereas \citet{Inoue_2020} used masked 5~GHz flux measured in \citet{Gallimore2004ApJ...613..794G}.
In the case of 1.4\,GHz, the upper limits after masking analysis are plotted.

\section{Discussion} \label{sec:discussion}
First, we investigate the beamsize issues for the representative SED obtained in Section\,\ref{sec:SED}.
Then, we investigate the possible origin of the SED considering single- and multi-components, respectively.

\subsection{beamsize issues} \label{sec:beam}
Since we restrict data sets with the angular resolution of $<0\farcs1$, we can safely ignore contamination from the host galaxy and compact radio emission components in the jet downstream such as components NE and C \citep{Gallimore_1996_1}. However, some components may have extended structures within the size of $\approx$0\farcs01--0\farcs1. In the following, we investigate the beamsize effects with our high-resolution data.

\subsubsection{Spectrum at $>10$\,GHz}
\begin{figure*}[!htbp]
\begin{center}
\includegraphics[angle=0,scale=0.9]{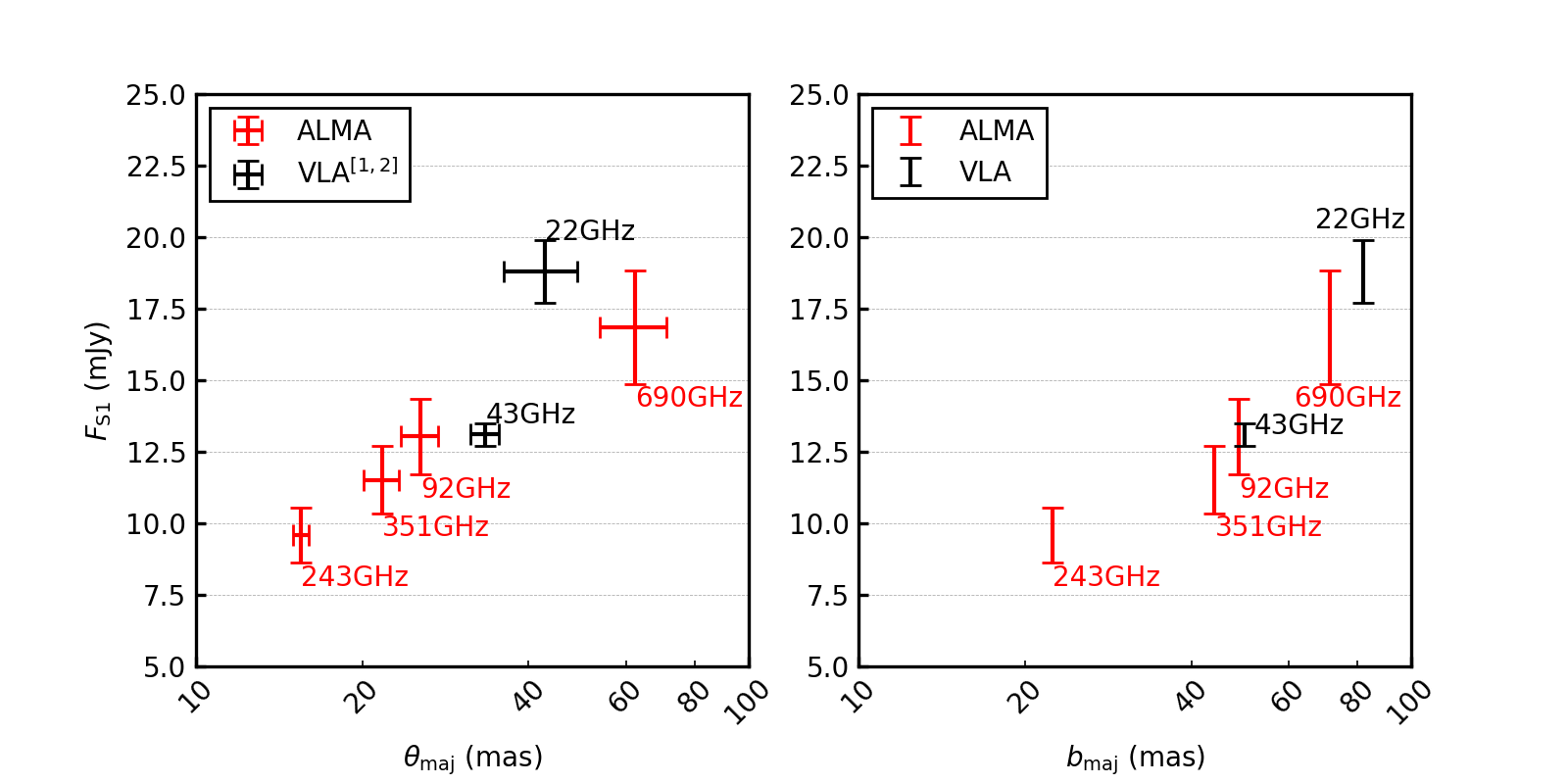}
\caption{\raggedright (left) The relation between beam-deconvolved source size and total flux (measured by 2D Gaussian fitting; {\tt imfit}) around S1. (right) We use a synthesized beam instead of the source size.
\label{fig:beam}}
\end{center} 
\end{figure*}

Figure~\ref{fig:rep_SED} and table~\ref{tab:imfit} show the flux rise/fall in $>10$\,GHz data. To check whether this fluctuation is due to beamsize issues or not, we investigated the relation between source size and flux at S1 measured by the 2D Gaussian fit (figure~\ref{fig:beam}).
The positive relation is confirmed in which the larger flux is seen in the larger source size. 
When we use the synthesized beamsize instead of the source size, the same positive trend is seen. 
This demonstrates that the flux decreasing trend from ALMA Band~6 to Band~9 seen in figure~\ref{fig:rep_SED} may be due to flat SED violated by beamsize effects, meaning that the flux rise/fall in $>10$\,GHz may not represent the physically motivated spectral index. 
In addition, figures~\ref{fig:rep_SED} and \ref{fig:beam} show that the $<100$\,GHz VLA data should be simultaneously considered to explain $>100$\,GHz ALMA data.
Therefore, we may need to reconsider the SSA emission suggested in \citet{Inoue_2020} and optically thick free-free emission suggested in \citet{Baskin2021MNRAS.508..680B} 
because both models rely on the flux rise from 256\,GHz to 694\,GHz which might be due to beamsize effects.

\subsubsection{Spectrum at $<10$\,GHz}
Understanding the spectrum $<10$\,GHz is also not straightforward.
At 8\,GHz, the flux of $5.4\pm0.5$~mJy was measured by VLBA observations \citep{Gallimore2004ApJ...613..794G}.
The drop from millimeter flux may not be real as suggested in \citet{Gallimore2004ApJ...613..794G} due to missing flux issues. 
For the 4.8\,GHz, VLA observations with large beamsize (i.e., $400$~mas) show the smaller flux than the 22\,GHz higher resolution data.
This suggests a marginally positive spectral index around 4.8\,GHz.
However, \citet{Gallimore2004ApJ...613..794G} suggest marginally negative (i.e., $s=-0.17$) index. Therefore, it is difficult to determine the spectral index at $<10$\,GHz, but it is not quite far from flat.
For VLBA observations, the robust information regardless of beamsize issue is the non-detection at 1.4\,GHz.
As shown in figure~\ref{fig:rep_SED}, the upper-limit flux at  1.4\,GHz is $<0.7$\,mJy while the achieved beamsize is larger and the uv-sampling is denser than 4.8\,GHz and 8\,GHz. Therefore, the significant flux drop between 4.8\,GHz to 1.4\,GHz is robust.

\subsection{Origin of the centimeter/submillimeter emission}
We investigate the origin of a centimeter to submillimeter broad band radio spectrum.
The SED can be explained by three components: Synchrotron emission from the jet base, grey-body emission from the dusty torus, and synchrotron emission from the X-raying corona (section \ref{sec:3comp}). However, we can not observationally rule out a single free-free scenario (section \ref{FF})  and a jet + dust scenario  (section \ref{sec:2comp}).
The details are explained in the following subsections.

\subsubsection{Single component (free-free)}\label{FF}
One possible radiation mechanism to explain the centimeter/submillimeter spectrum (figure~\ref{fig:rep_SED}) by a single flat-spectrum component (see also \cite{BH_position}) assuming that flux rise/fall in $>10$\,GHz is due to beamsize effects is free-free emission and self-absorption. 
The specific luminosity regarding free-free emissions at frequency $\nu$ is
\begin{equation}
\left[\frac{L_{\nu,\rm{ff, em}}}{\rm erg\,s^{-1}\,Hz^{-1}}\right] = \left[\frac{V}{\rm cm^{3}}\right]\times\left[\frac{\epsilon_\nu^{\mathrm{FF}}}{\rm erg\,cm^{-3}\,s^{-1}\,Hz^{-1}}\right],
\end{equation}
where the source unit volume ($V$) and $\epsilon_\nu^{\mathrm{FF}}$ is the emissivity of bremsstrahlung.
The emissivity is given by
\begin{eqnarray}
 \left[\frac{\epsilon_\nu^{\mathrm{FF}}}{\rm erg\,cm^{-3}\,s^{-1}\,Hz^{-1}}\right] &=&
6.8\times{10}^{-38}g\left(\nu,T_\mathrm{e}\right)
\times\left[\frac{T_\mathrm{e}}{\mathrm{K}}\right]^{-0.5}
\nonumber\\
&&
~~~\times\left[\frac{n_\mathrm{e}}{\mathrm{cm}^{-3}}\right]^{2}
\exp{\left(h\nu/k_\mathrm{B}T_\mathrm{e}\right)} 
\end{eqnarray}
\begin{equation}
g\left(\nu,T_\mathrm{e}\right)=0.5535\ln{\left|\left[\frac{T_\mathrm{e}}{\mathrm{K}}\right]^{1.5}\left[\frac{\nu}{\mathrm{GHz}}\right]^{-1}Z^{-1}\right|}-1.682,    
\end{equation}
where $T_\mathrm{e}$ is the temperature of the electron, $n_\mathrm{e}$ is the density of the electrons, and $g\left(\nu,T_\mathrm{e}\right)$ is the gaunt factor ($h$ and $k_{\rm B}$ are Planck’s constant and Boltzmann’s constant). 
Here, we approximate the emitting regions by uniform cylinders whose axis is the line of sight; i.e., $V = \pi (d_{\rm s}/2)^2\times l_{\rm L.O.S}$ where $d_{\rm s}$ is the diameter of emitting regions and $l_{\rm L.O.S}$ is the depth of the emitting regions along the line of sight (L.O.S).
The emission measure ($EM$) of the emitting region is defined by the integral of $n_{\rm e}^2$ along the line of sight
\begin{equation}
\left[\frac{EM}{\rm pc\,cm^{-6}}\right]\equiv
\int_{\rm L.O.S} n_\mathrm{e}^{2}{\rm d}l
= \left[\frac{n_\mathrm{e}}{\mathrm{cm}^{-3}}\right]^{2}
\left[\frac{l_{\rm L.O.S}}{\mathrm{pc}}\right].
\end{equation}
Free-free opacity is approximately provided by \citet{Mezger1967ApJ...147..471M} as
\begin{equation}\label{eq:tau}
\tau_{\nu}=3.28\times10^{-7}
\left[\frac{T_\mathrm{e}}{10^4\,\mathrm{K}}\right]^{-1.35}
\left[\frac{\nu}{\mathrm{GHz}}\right]^{-2.1}
\left[\frac{EM}{\mathrm{pc\,cm^{-6}}}\right].
\end{equation}
Considering free-free self-absorption, the free–free spectral luminosity is given by
\begin{equation}
\left[\frac{L_{\nu}}{\rm erg\,s^{-1}\,Hz^{-1}}\right] = 
\left[\frac{L_{\nu,\rm{ff, em}}}{\rm erg\,s^{-1}\,Hz^{-1}}\right]\exp({-\tau_{\nu}}).
\end{equation}

\begin{figure*}[!htbp]
\begin{center}
\includegraphics[angle=0,scale=0.24]{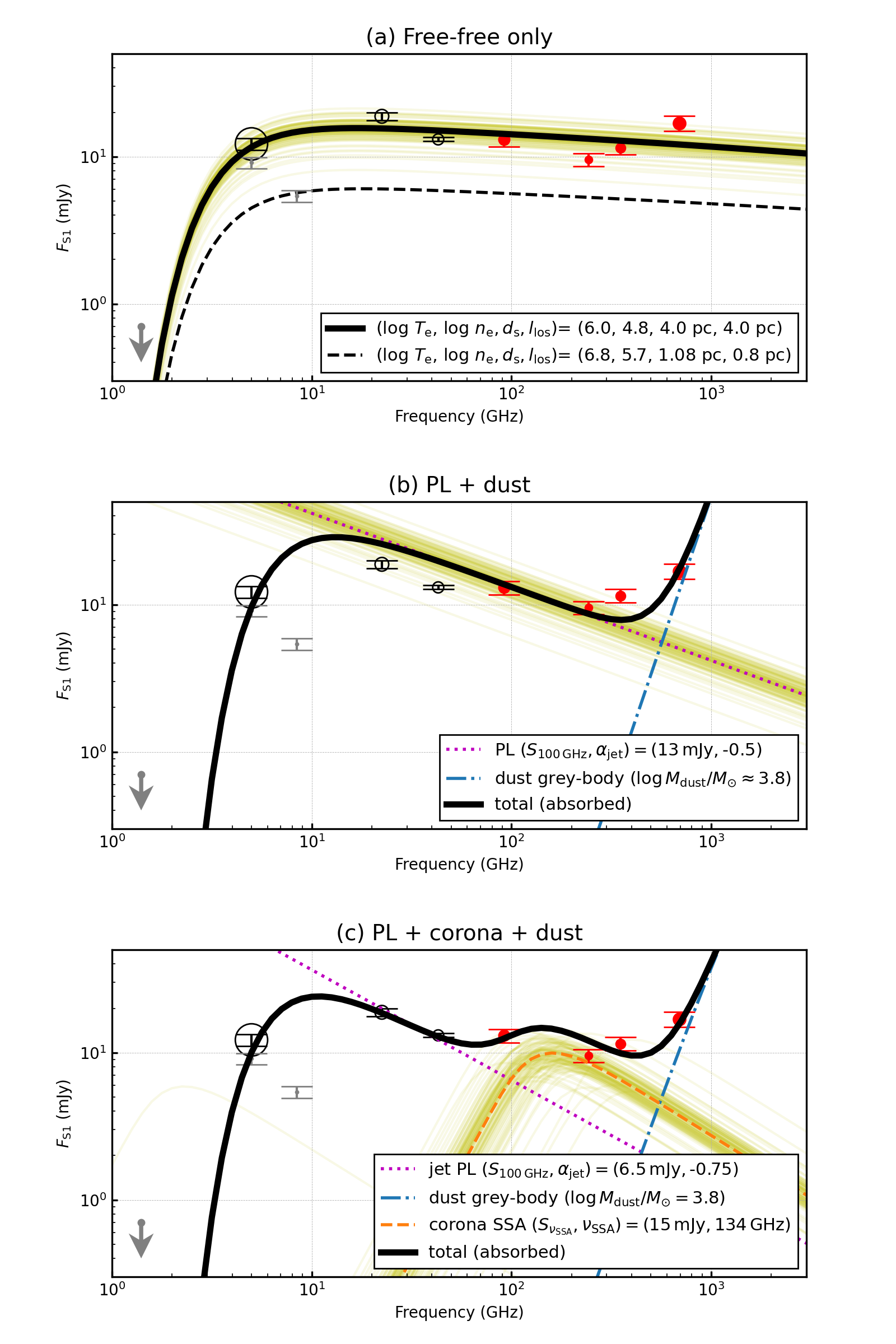}
\caption{\raggedright 
(a) The single-component model assuming free-free emission/absorption described in Section~\ref{FF}.
The black line is plotted based on Bayesian parameter estimation for $n_{\rm e}$. The black dashed line indicates the free-free emission and absorption with parameters shown in \citet{Gallimore2004ApJ...613..794G}.
(b) The two-components model using equations~(\ref{eq:PL}) and (\ref{eq:grey}). The red-dotted and blue dash-dotted lines are plotted based on Bayesian parameter estimation for $\alpha_{\rm 100GHz}$ and $M_{\rm dust}$.
(c) The three-components model explained by equation~(\ref{eq:total}). The purple dotted line is the synchrotron PL emission with $(S_\mathrm{100\,GHz},\alpha_{\rm jet})=(6.5\,\mathrm{mJy}, -0.75)$. The blue dash-dotted line indicates the dust grey-body with the dust mass of $M_\mathrm{dust}=6\times10^3\,M_{\odot}$. The orange dashed line indicates the SSA components originally from the corona ($S_{\nu_\mathrm{SSA}},\nu_\mathrm{SSA}$)=($15$\,mJy, 134\,GHz) which are estimated by Bayesian parameter estimation.
The yellow lines show the 100 samples from the chain during Bayesian parameter estimation.
\label{fig:all}}
\end{center} 
\end{figure*}

The equations in the previous paragraph show that we can determine the spectrum when we know ($n_\mathrm{e}$, $T_\mathrm{e}$, $d_\mathrm{s}$, $l_\mathrm{L.O.S}$).
According to \citet{Gallimore2004ApJ...613..794G}, the 8.4\,GHz VLBA flux and the rapid drop at 1.4\,GHz can be explained by the combination of parameters such as ($n_\mathrm{e}$, $T_\mathrm{e}$, $d_\mathrm{s}$, $l_\mathrm{L.O.S}$) = ($5\times10^5$~cm$^{-3}$, $6\times10^6$\,K, 1.1\,pc, 0.8\,pc) (figure~\ref{fig:all}a grey dotted line). 
Assuming the lower temperature and larger emitting region ($T_\mathrm{e}$, $d_\mathrm{s}$, $l_\mathrm{L.O.S}$) = ($10^6$\,K, 4\,pc, 4\,pc), our VLA and ALMA measurements indicate the electron density of $n_\mathrm{e}\approx6\times10^4$~cm$^{-3}$ based on Bayesian parameter estimation developed by \citet{Foreman-Mackey2013PASP..125..306F}\footnote{\url{https://emcee.readthedocs.io/en/v2.2.1/user/line/}}.
In this model, the inverted spectral feature at the Bands 6, 7, and 9 ranges in the SED can be explained by beam size effects (section~\ref{sec:beam}).

Understanding the origin of the ionized gas is non-trivial.
For example, the ionized gas temperature cannot be as high as the plasma where X-rays can be produced efficiently (i.e., $T_{\rm e}\approx10^8$~K at X-ray emitting plasma) because high temperature cannot explain the rapid drop at $<10$\,GHz. 
The relatively low-temperature clouds in the broad line region (BLR) can explain the SED. However, the typical BLR size is $<1000$ light days, i.e., $<1$\,pc \citep{Kaspi2000ApJ...533..631K}, which means that emission should be point-source and it is contradictory to the beamsize effects by spatially extended structure.
The gas compressed by the radiation pressure above the BLR would be the case. However, in that case, the SED drops above $>100$\,GHz according to the model by \citet{Baskin2021MNRAS.508..680B}, which cannot explain the observed 92\,GHz flux density.
As suggested in \citet{Gallimore2004ApJ...613..794G}, free-free emissions from dense disk winds heated by X-ray from AGN can be the possible origin. Because the size of disk wind should be $\approx$pc to explain the emission flux, the spatially resolved structure should be confirmed by future high-resolution observations if a single-component free-free emission/absorption scenario is true.

\subsubsection{Multiple components}
The centimeter/submillimeter SED can be explained by an ensemble of multiple components.
One radiation mechanism in centimeter (low-frequency) is the power-law (PL) synchrotron components from the base of the jet 
\begin{equation}\label{eq:PL}
S_\mathrm{\nu,PL}=S_\mathrm{100\,GHz}\left(\frac{\nu}{100\,\mathrm{GHz}} \right)^{\alpha_{\rm jet}},
\end{equation}
where $S_{\rm 100\,GHz}$ is the normalization at 100\,GHz and $\alpha_{\rm jet}$ is the spectral index ($\approx-0.75$ is often adapted as representative value; \cite{Condon2016era..book.....C}).

In submillimeter (high-frequency), dust grey body from the torus may contribute \citep{Garc2016ApJ...823L..12G}
\begin{equation}\label{eq:grey}
S_{\nu,\mathrm{dust}} = M_\mathrm{dust}\times\kappa_\nu\times B_\nu(T_\mathrm{dust})/D_\mathrm{L}^2,
\end{equation}
where $M_\mathrm{dust}$ is the dust mass, $T_\mathrm{dust}$ is the dust temperature, $\kappa_\nu$ is the dust emissivity
($\sim\kappa_\mathrm{352GHz}\times(\nu\,\mathrm{[GHz]}/352)^\beta$ 
with 
$\kappa_\mathrm{352GHz}=0.09\,\mathrm{m^2 kg^{-1}}$), 
$\beta$ is the emissivity index, and $D_\mathrm{L}$ 
is the luminosity distance.
We note that the infrared interferometer, GRAVITY, has revealed the existence of hot dust near the black hole \citep{GRAVITY2020A&A...634A...1G,BH_position}. However, this hot dust does not contribute to the ALMA high-frequency observation. ALMA high-frequency observations rather trace cold dust, whose property is still uncertain. Therefore, we assume $T_{\rm dust}=46$~K and $\beta=2$ according to \citet{Garc2016ApJ...823L..12G} for cold dust.

In millimeters, self-absorption (SSA) at the corona would be seen \citep{Inoue2014PASJ...66L...8I, Inoue2018ApJ...869..114I, Inoue2019ApJ...880...40I, Kawamuro2022ApJ...938...87K,Ricci2023arXiv230604679R}, which are characterized by
\begin{eqnarray}\label{eq:SSA}
S_\mathrm{\nu,SSA}=S_\mathrm{\nu_{SSA}}&&\left( \frac{\nu}{\nu_\mathrm{SSA}} \right)^{5/2}\\ \nonumber
&&\left\{1-\exp\left[\left(-\frac{\nu}{\nu_\mathrm{SSA}} \right)^{-(\delta+4)/2}\right]\right\}
\end{eqnarray}
where $S_\mathrm{SSA}$ is the normalizations, $\nu_\mathrm{SSA}$ is the SSA frequency, and $\delta$ is the slope of the electron-energy spectrum (the slope of $\alpha_{\rm corona}=(1-\delta)/2$ in the SED).
Determining $S_\mathrm{SSA}$ and $\nu_\mathrm{SSA}$ enables us to estimate the coronal magnetic field strength ($B$) and the size ($R$); e.g., $B\propto \nu_\mathrm{SSA} S_\mathrm{SSA}^{-0.1}$ and $R\propto \nu_\mathrm{SSA}^{-1} S_\mathrm{SSA}^{0.5}$ when $\delta=2.7$. The detailed formulation is shown in \citet{Chaty2011A&A...529A...3C}.

\subsubsection{Two components (jet + dust)}\label{sec:2comp}
Figure~\ref{fig:all}(b) tries to explain the SED by two components, i.e., jet PL (purple) and dust grey body (blue).
Even if we assume the PL index of $\alpha_{\rm jet}\approx-0.5$ (without energy loss at the high frequency), the VLA data points are systematically lower than the predicted line (black) and Band~6 and 7 points are higher.
These systematic offsets can not be explained by beamsize effects seen in figure~\ref{fig:beam}.
It should be noted that a flat jet spectra ($\alpha_{\rm jet}>-0.5$) may not be likely in the case of the nucleus of NGC~1068. Such a flat spectrum appears in blazars having strong beaming effects \citep{Itoh2020ApJ...901....3I}, however, this is not the case in NGC~1068. In an extremely compact jet case, a flat spectrum could appear even in Seyferts \citep{Anderson2004ApJ...603...42A,Falcke2004A&A...414..895F}. However, this scenario is also unlikely because the VLBI 8.4~GHz flux is much smaller than the VLA flux under the assumption of a flat spectrum.
Therefore, two components scenario might be unlikely and the additional components such as coronal SSA may be necessary to explain the SED seen in figure~\ref{fig:rep_SED}.

\subsubsection{Three components (jet + corona + dust) scenario}\label{sec:3comp}
\citet{Inoue_2020} have argued SSA components originally from the corona, however, it is necessary to revisit the model because our new 100\,GHz data point cannot be reproduced by their model.
Figure~\ref{fig:all}(c) demonstrates that the three-components scenario can explain the SED at S1 in NGC~1068.
While the VLA flux is considered as the upper limit in \citet{Inoue_2020}, we account for the 22\,GHz flux given the new 100~GHz data point.
We consider that the 22\,GHz continuum is mainly from synchrotron emission with $(S_\mathrm{100\,GHz},\alpha_{\rm jet})=(6.5\,\mathrm{mJy}, -0.75)$ (purple dashed line in figure~\ref{fig:all}c).
This may explain the possible polarization detection at 22\,GHz \citep{Gallimore_1996_2}. 
For the high frequency (i.e., ALMA Band~9), we consider the grey-body with the dust mass of $M_\mathrm{dust}=6.5\times10^3\,M_{\odot}$ (blue dashed line in figure~\ref{fig:all}).
Because the PL at low frequency and grey-body at high frequency cannot explain the excess of 90-400\,GHz emission, we consider the SSA components originally from the corona.
Assuming $\delta=2.7$ in equation~(\ref{eq:SSA}), the best fit parameter is
($S_{\nu_\mathrm{SSA}},\nu_\mathrm{SSA}$)$=$($15^{+3}_{-3}$\,mJy, $134^{+28}_{-22}$\,GHz)
based on Bayesian parameter estimation (the orange dashed line in figure~\ref{fig:all}c).
The SED fit does not reproduce the inverse spectrum between Band~6 and 7. However, this can be easily reconciled by considering the spectral softening of the jet around Band~6. Compared to figure~1 of \citet{Inoue_2020}, our new analysis improves the overall behavior of the SED by including new 100~GHz data.
The brightness temperature is estimated to be $410\pm80$\,K at $\nu_\mathrm{SSA}$ considering the synthesized beam size of 0\farcs05.
According to the coronal model \citep{Inoue2018ApJ...869..114I}, the derived $S_{\nu_\mathrm{SSA}}$ and $\nu_\mathrm{SSA}$ indicate the coronal magnetic field strength of 
$B=16^{+5}_{-4}$\,G on scales of $R=29^{+6}_{-6}$ Schwarzschild radii 
from the central black holes. When we estimate $B$ and $R$, we adopt the energy fraction of non-thermal electrons of 0.03, the coronal temperature of $10$\,keV \citep{Pal2022MNRAS.517.3341P}, and the Thomson scattering optical depth value of 1.1, and the black hole mass of $5\times10^7\,M_{\odot}$ according to \citet{Inoue_2020}.
The error terms of $S_{\nu_\mathrm{SSA}}$, $\nu_\mathrm{SSA}$, $B$, and $R$ are calculated based on the 16th, 50th, and 84th percentiles of the samples in the marginalized distributions. The systematic errors regarding assumptions (e.g., several coronal parameters such as a typical energy fraction of non-thermal electrons and the Thomson scattering optical depth) are not considered.
Compared to the previous investigation by \citet{Inoue_2020}, $\nu_\mathrm{SSA}$ is at the lower frequency in order to reconcile the new 100\,GHz data point. This is because we have a smaller coronal magnetic field value than that reported in \citet{Inoue_2020} ($B\approx100$\,G). We note that our new magnetic field estimate $B\approx20$\,G is consistent with the values revealed in other nearby Seyferts \citep{Inoue2018ApJ...869..114I}.

\subsubsection{Screen by ionized gas}
In figures~\ref{fig:all}b and \ref{fig:all}c, to explain the rapid drop at $<10$\,GHz, we consider that the emissions are screened by free–free absorption originally from spatially extended diffuse ionized gas components (which would be associated with the host galaxy) as suggested by \citet{Baskin2021MNRAS.508..680B}.
The black line in figure~\ref{fig:all}c corresponds
\begin{equation}\label{eq:total}
\left[\frac{L_{\nu}}{\rm erg\,s^{-1}\,Hz^{-1}}\right] = 
\left[\frac{L_{\nu,\rm{pl}}+L_{\nu,\rm{SSA}}+L_{\nu,\rm{dust}}}{\rm erg\,s^{-1}\,Hz^{-1}}\right]\exp({-\tau_{\nu}}).
\end{equation}
A possible interpretation of the origin of the absorber is diffuse ionized gas with the parameters of ($n_\mathrm{e}$, $T_\mathrm{e}$, $l_\mathrm{L.O.S}$) = ($3\times10^3$\,cm$^{-3}$, 
$10,000$\,K, $20$\,pc) which is typycal H\,\emissiontype{II} region in galaxies \citep{Hunt2009A&A...507.1327H}. However, investigating the origin of the absorbers is beyond the scope of this paper because the parameters of the absorber can not be determined uniquely.

\section{Summary}
We report that the nucleus region of Seyfert galaxy NGC~1068 has the flux density at 5--700\,GHz (centimeter/submillimeter) of $\approx$ 10--20\,mJy, which seems flat SED but determining the spectral index is non-trivial due to probable beamsize issues.
In particular, we found that previous models (e.g., \cite{Inoue_2020, Baskin2021MNRAS.508..680B}) were unable to account for the newly included 100\,GHz flux. Therefore, we undertook a reassessment of the emission mechanisms and made essential parameter adjustments.
One possible scenario which can explain the centimeter/submillimeter SED is an ensemble of multiple components (PL synchrotron components from the base of the jet, dust grey-body from the torus, and SSA from the corona).
This three-component scenario suggests the coronal magnetic field strength of $\approx20$\,G on scales of $\approx30$ Schwarzschild radii from the central black holes.
However, due to limited angular resolution, a single free-free and a jet + dust scenario are also possible scenarios.
Future ``high resolution" images obtained by e.g., ngVLA are necessary to distinguish spatially extended (i.e., jet and/or dust) and compact (i.e., corona) components. 

\begin{ack}
T.M. and Y.I. appreciate the support from NAOJ ALMA Scientific Research Grant Number 2021-17A.
T.M. is supported by JSPS KAKENHI grant No. 22K14073. Y.I. is supported by JSPS KAKENHI Grant Number JP18H05458, JP19K14772, and JP22K18277. This work was supported by World Premier International Research Center Initiative (WPI), MEXT, Japan.  
\end{ack}

\bibliographystyle{aa}
\bibliography{N1068_center}

\end{document}